\documentclass[]{mnras}
\usepackage[utf8x]{inputenc}
\usepackage{float}
\usepackage{graphicx}
\usepackage{amssymb}
\usepackage{amsmath}
\usepackage[usenames,dvipsnames]{xcolor}
\usepackage{multicol}
\usepackage{multirow}
\usepackage{enumerate}
\usepackage{natbib}
\usepackage{mathtools}
\usepackage{hyperref}

\DeclarePairedDelimiter\abs{\lvert}{\rvert}%
\DeclarePairedDelimiter\norm{\lVert}{\rVert}%

\makeatletter
\let\oldabs\abs
\def\abs{\@ifstar{\oldabs}{\oldabs*}}
\let\oldnorm\norm
\def\norm{\@ifstar{\oldnorm}{\oldnorm*}}
\makeatother

\newcommand{\msun}{M_\odot}

\hypersetup{draft}
\begin{document}

\raggedbottom	

\title[A Smoother Lens]{Through a Smoother Lens: An expected absence of LCDM substructure detections from hydrodynamic and dark matter only simulations}
	
\author[Graus et al.]{Andrew S. Graus$^1$\thanks{$\!\!$e-mail: agraus@uci.edu}, 
		James S. Bullock$^1$, Michael Boylan-Kolchin$^2$\\ \\\LARGE\rm Anna M. Nierenberg$^1$\\ \\
		$\!\!^1$Center for Cosmology, Department of Physics and Astronomy, 
		4129 Reines Hall, University of California Irvine, CA 92697, USA \\
		$\!\!^2$The University of Texas at Austin, Department of Astronomy, 2515 Speedway, Stop C1400, Austin, Texas 78712-1205\\}
\maketitle

\begin{abstract}
A fundamental prediction of the cold dark matter cosmology is the existence of a large number of dark subhalos around galaxies, most of which should be entirely devoid of stars.  Confirming the existence of dark substructures  stands among the most important empirical challenges in modern cosmology: if they are found and quantified with the mass spectrum expected, then this would close the door on a vast array of competing theories.  But in order for observational programs of this kind to reach fruition, we need robust predictions.  Here we explore substructure predictions for lensing using galaxy lens-like hosts at $z=0.2$ from the Illustris simulations both in full hydrodynamics and dark matter only.
We quantify substructures more massive than $\sim 10^9$ M$_\odot$, comparable to current lensing detections derived from HST, Keck, and ALMA. The addition of full hydrodynamics reduces the overall subhalo mass function by about a factor of two. Even for the dark matter only runs, most ($\sim 85\%$) lines of sight through projected cylinders of size close to an Einstein radius contain no substructures larger than $10^{9}$ M$_{\odot}$.  The fraction of empty sight lines rises to $\sim 95\%$ in full physics simulations.  This suggests we will likely need hundreds of strong lensing systems suitable for substructure studies, as well as predictions that include the effects of baryon physics on substructure, to properly constrain cosmological models.  Fortunately, the field is poised to fulfill these requirements. 
\end{abstract}

\section{Introduction} \label{s:intro}
	
One key prediction of the current paradigm of galaxy formation is that there should be some dark matter halo mass below which galaxies are unable to form \citep{Klypin99,Moore99,Bullock17}.  For example, if the (cold) dark matter particle is a neutralino,  then substructure should exist down to masses of order $10^{-6}$ $\msun$, depending on details of the model \citep{Green2004}. The halo mass at which galaxy formation is cut off should be many orders of magnitude higher than this. The specific cutoff scale depends on the complicated interplay between gas cooling and heating by an ultraviolet background but current estimates suggest that the mass scale is between $10^8$ and $10^9 $ $M_{\odot}$ \citep{Efstathiou92,Bullock00,Bovill09,Sawala2016a,Onorbe2016}. The implication is that if Cold Dark Matter (CDM) is the correct model, there should be numerous small dark matter halos with no galactic counterparts. As there is currently no detection of a dark matter particle the presence of such dark halos is a great way to constrain dark matter models.

Low-mass dark halos that exist within the virial radius of larger halos are known as subhalos and many techniques have been proposed to detect them.  These include searches for gaps or other features in stellar streams in the Galactic halo \citep{Johnston2002,Ibata2002,Carlberg09,Bovy2016}, and via the detection of products in the annihilation of dark matter particles into standard model particles \citep{Kuhlen08,Ng2014,Sanchez-Conde2011}. One additional promising avenue for detecting dark substructure is from strong lensing \citep{Mao98,Metcalf01,Dalal02}.

The detection dark substructure via lensing is in principle straightforward.  First, a suitably lensed object is detected, the foreground lens is modeled with a smooth potential for the host galaxy, and then perturbations are added to the foreground lens model in the form of potentials of possible dark matter substructure. Several systems have been analyzed with the hopes of detecting a subhalo, resulting in the detection of dark matter substructure at the $\sim 10^9$ M$_\odot$ mass scale for galaxy-galaxy lensing \citep{Vegetti2012,Hezaveh2016}, and $\sim 10^{7.5}$ M$_\odot$ from quasar flux ratio anomalies \citep{Nierenberg14}. This has allowed the authors to put constraints on the subhalo mass function and the fraction of mass that is in substructure. With current technology these methods could push down even further to M$_{600}$ $\sim 10^{7}$ M$_\odot$, however no structures that small have been detected \citep{Nierenberg17}. 

The field of substructure lensing is intriguing because there is great potential in the near future for a substantial increase in the number of lenses and the ability to detect smaller objects.  The Dark Energy Survey (DES), LSST,  Euclid, and WFIRST will lead to an enormous increase in the number of lensing systems appropriate for looking for dark substructure. As an example, \cite{Collett2015} estimates that DES, LSST and EUCLID will potentially discover 2400, 120,000 and 170,000 galaxy-galaxy lensing systems respectively.  Currently, ALMA has the capability to detect very small substructure, potentially probing subhalo masses of $10^{6}$ $\msun$ \citep{Hezaveh2016}. Furthermore, JWST will allow for detections at ($\simeq 10^{7}$ $\msun$) in halo mass, based on quasar flux ratio anomalies \citep{MacLeod13}. However, without accurate predictions for what is expected within CDM, these observational constraints will never reach their scientific fruition.

Up until just a few years ago, simulating large numbers of galaxies with full hydrodynamics at high resolution was not possible. Therefore, theoretical studies of substructure lensing have used  dark-matter-only simulations as their benchmark for comparison \citep[e.g.,][]{Vegetti2014}.  Recent simulations of Milky Way type galaxies  have shown that full hydrodynamic simulations produce significantly fewer bound substructures than their dark matter only counterparts \citep{Brooks14,Wetzel2016,Zhu2016,Sawala2016c,SGK2017}, with a factor of $\sim 2$ reduction within the virial radius and an order of magnitude fewer subhalos within $\sim 20$ kpc of the central galaxy. 
 The primary cause appears to be the enhanced central potential created by the host galaxy: subhalos get destroyed as they approach the central galaxy on radial orbits.  While this result is potentially of interest for subhalo lensing, most of the highest resolution simulations have been run at the Milky Way mass scale ($\sim$ $10^{12}$ $M_{\odot}$), as opposed to the mass scale used for substructure lensing studies which is closer to halo masses of $10^{13}$ $M_{\odot}$.   Fortunately, such simulations are just becoming feasible both in zoom-in simualtions \citep{Fiacconi2016} and in full box simulations, such as the EAGLE simulation \citep{Schaye2015}, and the Illustris simulation \citep{Vogelsberger2014}. This allows for studies of lensing substructures in simulations with full hydrodynamics. One example of this is \citet{Despali2016} who use the EAGLE simulation (both dark matter only and full hydrodynamics), and Illustris hydrodynamics simulation to investigate predictions for subhalo lensing. They find that hydrodynamic simulations decrease the average expected substructure mass fractions, as would be expected form results on the Milky Way scale. 
 
 The present paper expands upon this past work by presenting the distributions of substructure mass fractions seen in projection along many lines-of-sight to host halos in the Illustris and Illustris-Dark simulations.  We compare to the distribution inferred from the \citet{Vegetti2014} study as well projected substructure mass functions derived from \citet{Hezaveh2016}.  We explore how the distributions change as we go from dark matter only to full hydrodynamics to illustrate both average differences between the two and  and impact of halo-to-halo scatter on expected substructure detection probabilities.
 
The format of the paper is as follows.  Section 2 details the Illustris simulations, our host sample, and subhalo properties.   Section 3 shows our results on substructure mass fractions  and subhalo mass functions. Section 4 explores the implications for substructure lensing. 
	
\section{Simulations and Methods} \label{s:data}
	
We make use of the publicly available Illustris simulations \citep{Nelson2015,Vogelsberger2014}
for this work and adopt their cosmological parameters: $\Omega_m$ = 0.2726, $\Omega_\Lambda$ = 0.7274, $\Omega_b$ = 0.0456, $\sigma_8$ = 0.809, $n_s$ = 0.963, and $\it{h}$ = 0.704.  The Illustris suite consists of three hydrodynamic and three dark matter only (DMO) simulations of increasing resolution all initialized at $\it{z}$ = 127 in a box that is 106.5 Mpc (co-moving) on a side.  
We use only the highest resolution version,  Illustris-1, which has a dark matter particle mass of $m_{dm}$ = 6.3 $\times$ $10^6$ $M_{\odot}$ and a gas particle mass of $m_{gas}$ = 1.3 $\times$ $10^6$ $M_{\odot}$.  The DMO version of Illustris-1 subsumes the baryonic matter into the dark matter particles and thus has a DM particle mass $m_{dm}$ = 7.6 $\times$ $10^6$ $M_{\odot}$.  In what follows, when we compare dark matter halo masses from Illustris-1 to those in Illustris-1 Dark we account for the excess baryonic mass in the DMO run by multiplying halo masses by $(1-f_b)$, where $f_b \equiv \Omega_b/\Omega_m=0.167$.  This approach assumes that the dark matter (sub)halos of interest are effectively depleted of their baryons.  For the sake of brevity, we refer to the Illustris-1 and Illustris-1 Dark simulations  as simply ``Illustris"  and ``Illustris Dark" below.

\begin{figure*}
	\includegraphics[width=3.25in, height=3.25 in, trim = 1.5in 0 1.0in 0]{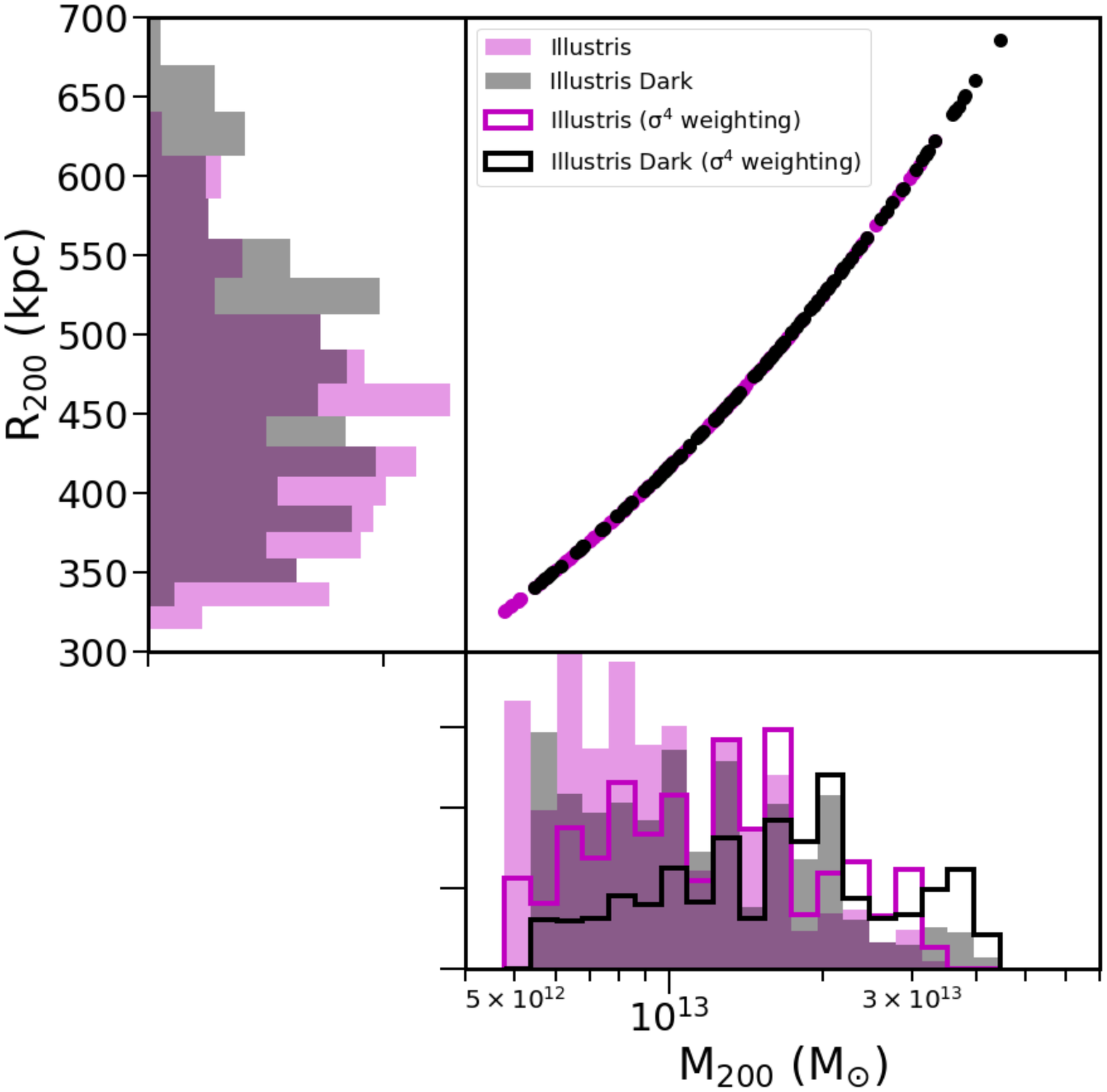}
	\hspace{.1in}
	\includegraphics[width=3.25 in, height=3.25 in, trim = 1.5in 0 1.0in 0]{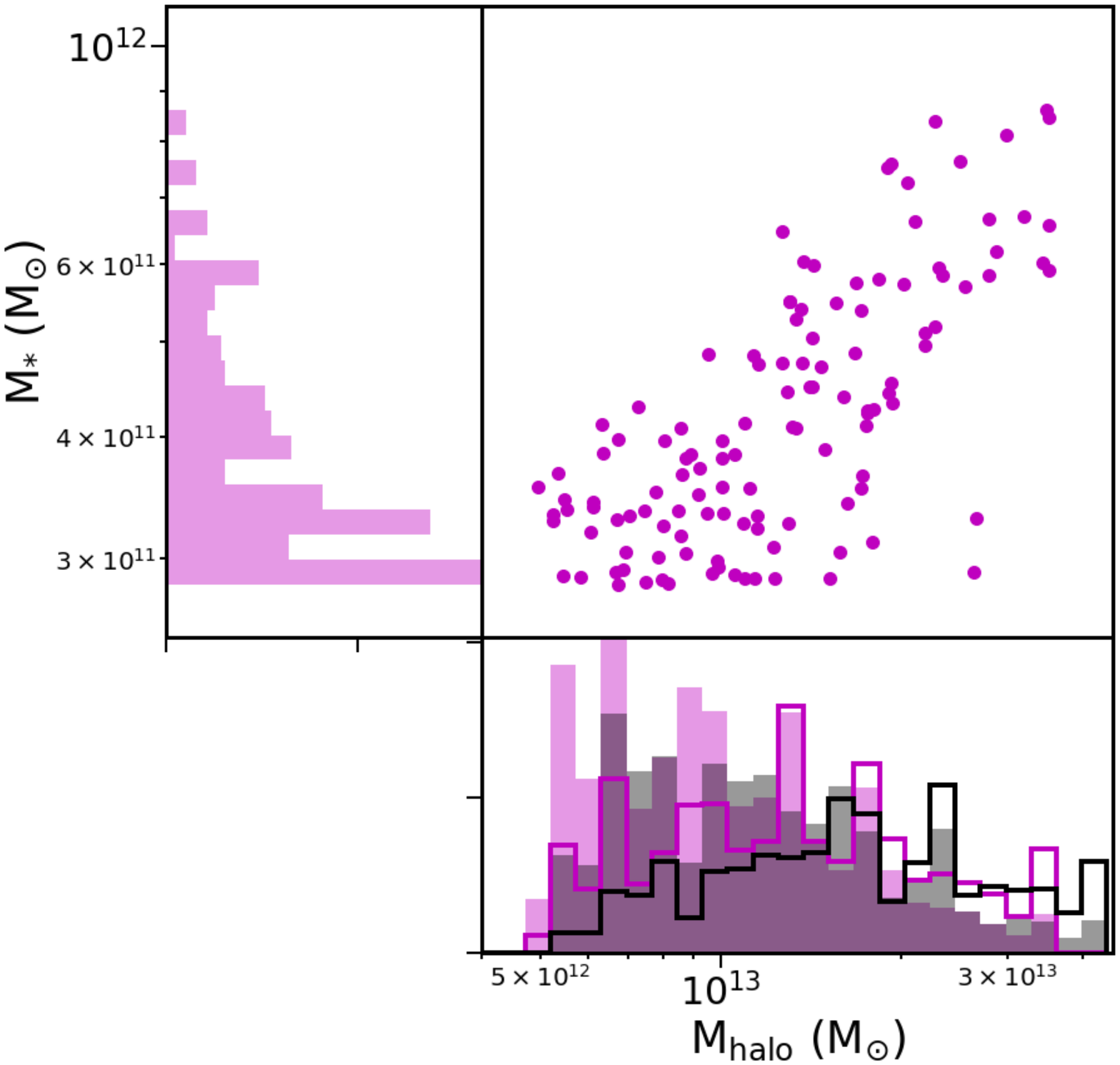}
	\centering
	\caption[Host Halo properties]{Properties of the host halos of our mock lens sample taken from Illustris and Illustris Dark selected such that they are isolated (as described in text) and that they are identified as the same halo in both Illustris and Illustris Dark.   Left: $M_{200}$ versus $R_{200}$ for the sample. Right: $M_{200}$ versus galaxy stellar mass $M_{\star}$ (for Illustris). The shaded histograms represent the fiducial sample, and the unfilled histograms represent the results of reweighting the systems by $\sigma^{4}$, reflecting the fact that systems for which a multiply lensed galaxy is detected are biased towards larger halos (see Section 2).} 
	\label{fig:one}
\end{figure*}

 \subsection{Host halo selection}
Dark matter halos in the Illustris catalogs were derived using a friends-of-friends (FoF) algorithm with a linking length of 0.2 times the mean inter-particle separation. After these FoF halos were identified, SUBFIND \citep{Springel01} was run on them in order to identify substructures around the host halos. For the rest of this work, we rely on two of the halo mass measurements provided in the Illustris data release. The first of these masses is $M_{200}$, which is defined as the total mass enclosed in a sphere of radius $R_{200}$ that has a mean density 200 times the critical density of the universe. The second mass parameter, which we refer to as $M_{\rm halo}$, is the mass that is bound to that halo but not bound to any of its subhalos. We will also refer to halos and subhalos in terms of their $V_{\rm max}$, which is the maximum circular velocity of the halo.  In our substructure analysis, we consider every subhalo within $R_{200}$ of the host and usually consider only subhalos with $M_{\rm halo} > 10^{9}$ $M_{\odot}$ (with more than 130 particles) in order to ensure completeness. 

Our aim is to explore substructure in massive halos of the type associated with strong-lens systems focusing on the distribution of mass fractions, and substructure mass functions.  As a benchmark, we refer to the sample used in \cite{Vegetti2014}, which has  an average redshift of of $\langle z \rangle$ = 0.233  and consists of galaxies with stellar masses $M_{\star} = 10^{11.45 - 11.95}$ $M_{\odot}$.  
Motivated by this, we focus our study on galaxy halos in the the $z = 0.2$ Illustris snapshot that host galaxies in this stellar mass range. We also make sure that the subhalo containing the galaxy is the most massive system in its FoF group, this works as a rough isolation criteria.  We then create a matched DMO sample by using matching files provided in the public Illustris data release to match to Illustris Dark.  This leaves us with a lens host sample of 122 halos in both Illustris and Illustris Dark.   Figure \ref{fig:one} shows the basic properties of our host galaxy-halo sample, with the dark matter only halo masses corrected by the baryon fraction (1-$f_{b}$). Interestingly, even after the hosts from Illustris Dark are corrected for the baryon fraction, they still end up being slightly more massive than the matched sample from Illustris.  This may be associated with the dynamical loss of dark matter in response to explosive feedback episodes \citep{Taylor16}.

Our sample of simulated galaxies covers the same stellar mass as the \citet{Vegetti2014} lens sample, with a $M_{200}$ range of $5\times10^{12}$ to $4\times10^{13}$ $M_{\odot}$. In terms of $V_{\rm max}$, this translates to halos between 300 and 700 km $\rm s^{-1}$. Note that while our sample has been selected from the same stellar mass range as the \citet{Vegetti2014} sample, it has a different distribution. This is because while there are more galaxies at smaller masses, galaxies at larger masses are more likely to be strong lenses. In order to account for this, we weight the distribution of galaxies by $\sigma^{4}$, where $\sigma$ is the velocity dispersion of halo. This is roughly how lensing strength scales with increasing mass \citep{Bourassa1973}, and once correcting for this effect, our distribution of stellar masses looks similar to that from the \citet{Vegetti2014} sample. The resulting  distribution is shown in Figure \ref{fig:one}. For the rest of this work, we will weigh our samples by the same factor.

\begin{figure*}
		\includegraphics[width=\textwidth, height=4.0 in, trim = .2in 1.3in 0 1.3in]{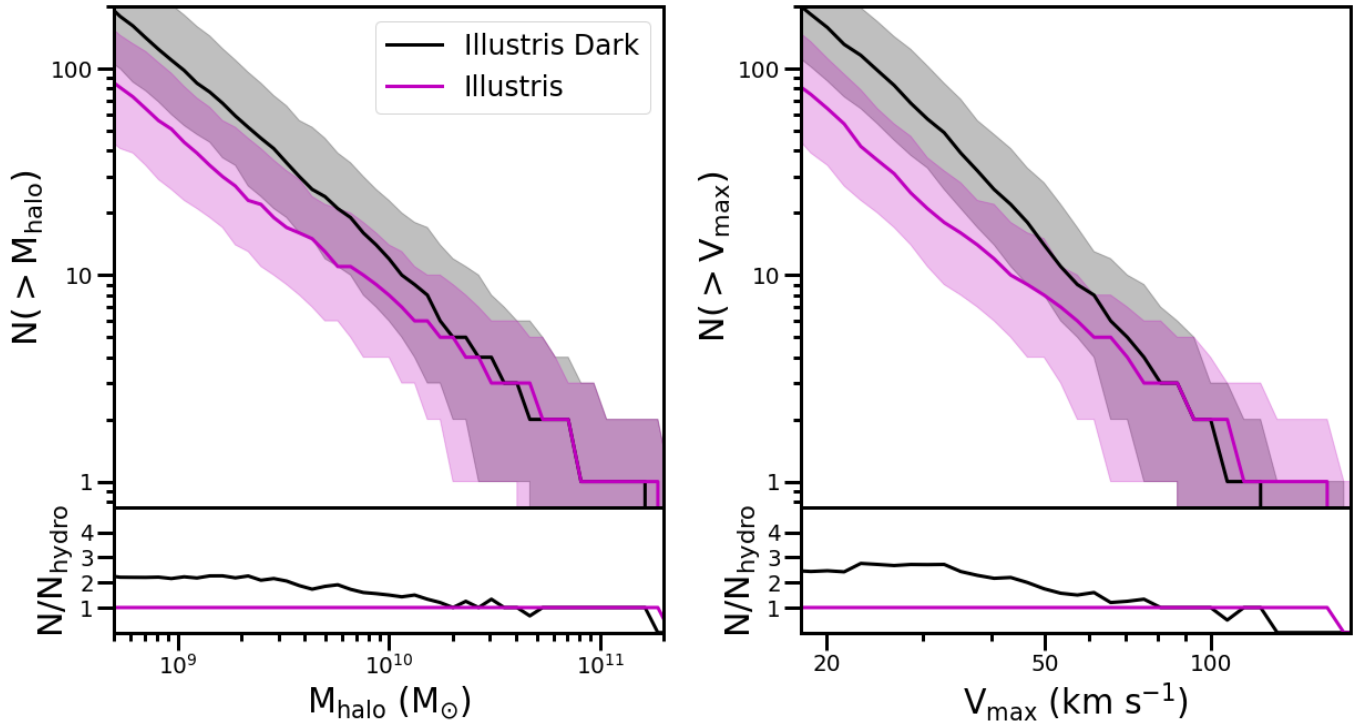}
		\centering
		\caption[HMFandVMF]{Median subhalo mass functions (left) and $V_{\rm max}$ functions (right) for our host galaxy samples from Illustris (magenta) and Illustris Dark (black).  Lines show medians, with $68\%$ shown as shaded bands.  As emphasized in the lower panels, the DMO runs contain roughly twice as many subhalos as the hydrodynamic runs.}
		\label{fig:hmf}
\end{figure*}

\subsection{Subhalo selection}

For the substructure, we restrict ourselves to the Illustris subhalo catalogs. Furthermore, we restrict ourselves to halos within 2 $\times$ $R_{200}$ of the host halo. It is important to note that small halos along the line of sight to a lens could also cause perturbations and also influence the lensing signal \citep{Metcalf05,Despali17}.

One important issue concerns the definition of subhalo mass. There are at least two separate classes of mass ambiguity important for substructure lensing studies. The first has to do with translating a reported subhalo mass from a lensing analysis to a subhalo mass as measured in a simulation.  The second has to do with how one defines subhalo mass in a simulation from the outset.  We refer the reader to \cite{Minor2016} for a discussion of how the shape of the assumed perturber's density profile affects strongly the mapping between the mass reported for a subhalo in simulations and the mass usually assumed in lensing models. The second issue is explored in Appendix \ref{s:AppA}, where we show that different halo finders can lead to a factor of $\sim 2$ differences in subhalo mass functions derived for the same simulation.  As we move towards an era where strong lenses suitable for substructure constraints become more common, the community will need to agree upon the best way to characterize subhalos and their masses to make progress in constraining the nature of dark matter or galaxy formation on small scales.  In this paper we adopt the subhalo masses as defined in the primary Illustris catalog.

We note that the Illustris catalogs contain a few low-mass items in the subhalos catalogs (about one per host) that are fully baryon-dominated (with more than 50\% of the total mass baryonic).  These unusual systems are typically found close to the central galaxy (see Appendix \ref{s:AppB}) and are potentially associated with SUBFIND identifying part of the host galaxy as a separate galaxy subhalo. In general, these baryon-dominated halos tend to be star particle dominated at very small radii (r $<$ 10 kpc) and gas dominated further out (10 kpc $<$ r $<$ 25 kpc). For the rest of this analysis, we throw out any baryon-dominated subhalos because we are interested only in prospects for detecting actual dark matter substructure.  However, it is worth noting that any real clumps of gas or stars (e.g. star clusters, tidal dwarf galaxies, or gas clouds) in the vicinity of lens hosts could be a significant background in lensing searches for dark substructure. Studies of this effect have estimated the contamination to be about 10\% although the specifics depend on the lens method, and the mass of the perturber \citep{Hsueh2017b,Gilman2017,He17}.

\section{Results} \label{s:discussion}

The left panel of Figure \ref{fig:hmf} shows subhalo mass functions (within $R_{200}$) for our  Illustris (magenta) and Illustris Dark (black) host samples.  The lines show median values and shaded bands show 68 percentile confidence intervals.   Subhalo masses in Illustris Dark have been scaled  $M_{\rm halo} \rightarrow (1-f_b) M_{\rm halo}$ in order to account for the baryonic mass subsumed by the `dark matter' particles in the DMO run.    The right panel of Figure \ref{fig:hmf} shows the corresponding  subhalo  $V_{\rm max}$ functions (again with $V_{\rm max} \rightarrow \sqrt{1-f_b} V_{\rm max}$ in the DMO runs).   The full physics runs display fewer subhlaos in both presentations.  The bottom panels quantify the difference by showing the ratio of subhalo counts in the DMO simulations to the hydrodynamic simulations.  The hydrodynamic simulations have about a factor of two fewer subhalos at fixed subhalo mass than the DMO simulations, even after accounting for baryonic mass loss.  

\begin{figure}
	\includegraphics[width=3.5in, height=3.0 in, trim = 1in 0 0 0]{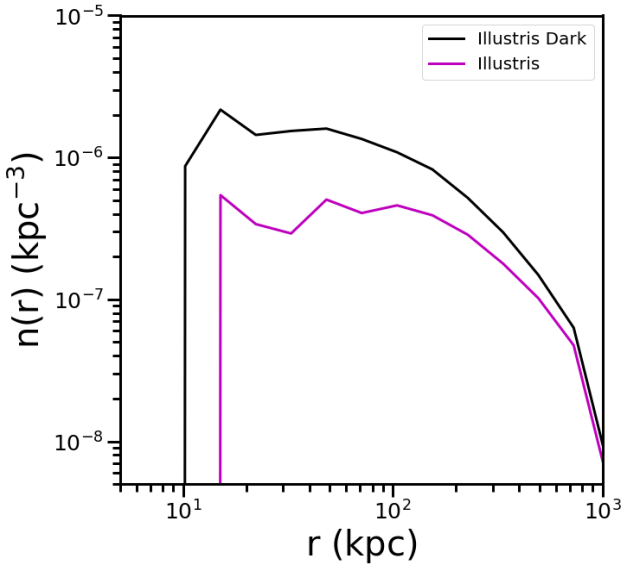}
	\centering
	\caption[Radial distribution]{Average 3D radial distributions of dark subhalos more massive than $M_{\rm halo} = 1\times10^{9}$ $M_{\odot}$ in Illustris (magenta) and Illustris Dark (black) as a function of radius $r$.  Shown are averages over all host halos in our large sample of hosts. Not only is the overall count of subhalos reduced in the hydro runs, but their radial distribution is less centrally concentrated.  This strongly suggests that dynamical disruption of substructure from the central host galaxy plays a major role reducing substructure counts compared to the DMO simulation.}
	\label{fig:radial}
\end{figure}

What is the cause of this substructure depletion?  As seen in previous work focusing on Milky Way size halos \citep[]{Brooks14,Wetzel2016,Zhu2016,Sawala2016c,SGK2017}, the origin appears to be subhalo interactions with the central galaxy.
Evidence for this is presented in Figure \ref{fig:radial}, which shows the average differential number density of subhalos as a function of radius in the DMO (black) and full physics (magenta) samples.  Here we include all subhalos more massive than $M_{\rm halo} = 10^9 \msun$.  We see that not only are there fewer subhalos in Illustris compared to Illustris Dark, but the radial distribution is significantly depleted in the central region.  There is a sharp cutoff in dark subhalo counts within $r \simeq 20$ kpc, which we attribute to interactions with the central host galaxy. This lack of expected dark subhalos at small radius could potentially be important for substructure lensing studies, which are sensitive to subhalos at small projected radius, comparable to the host lens Einstein radius (typically $\lesssim 10$ kpc).  We explore implications for lensing explicitly in the next subsection.

\begin{figure*}
	\includegraphics[width=3.2in, height=3.2 in, trim = 0.5in 0 0.5in 0]{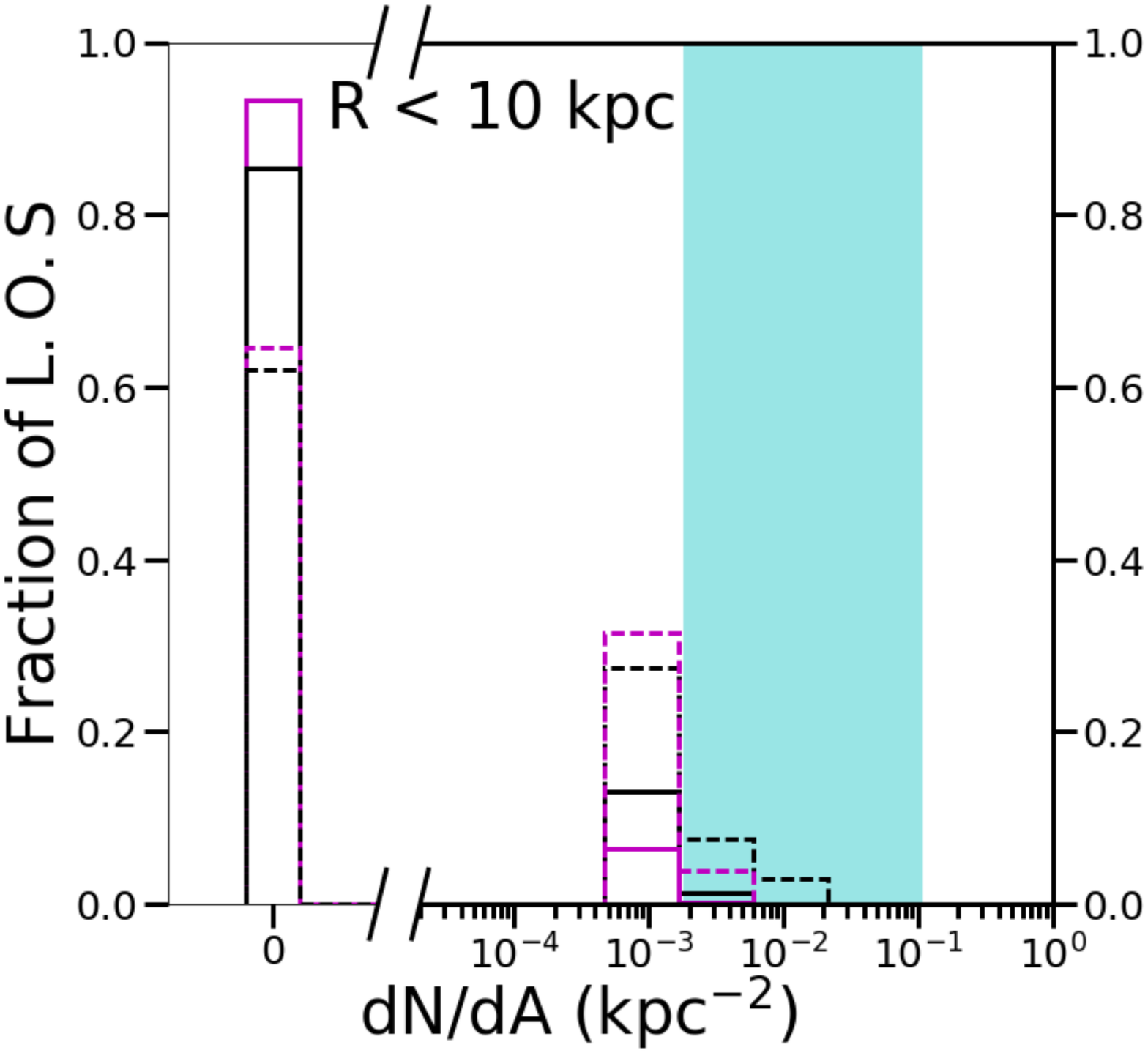}
	\hspace{.2in}
	\includegraphics[width=3.2 in, height=3.2 in, trim = 0.5in 0 0.5in 0]{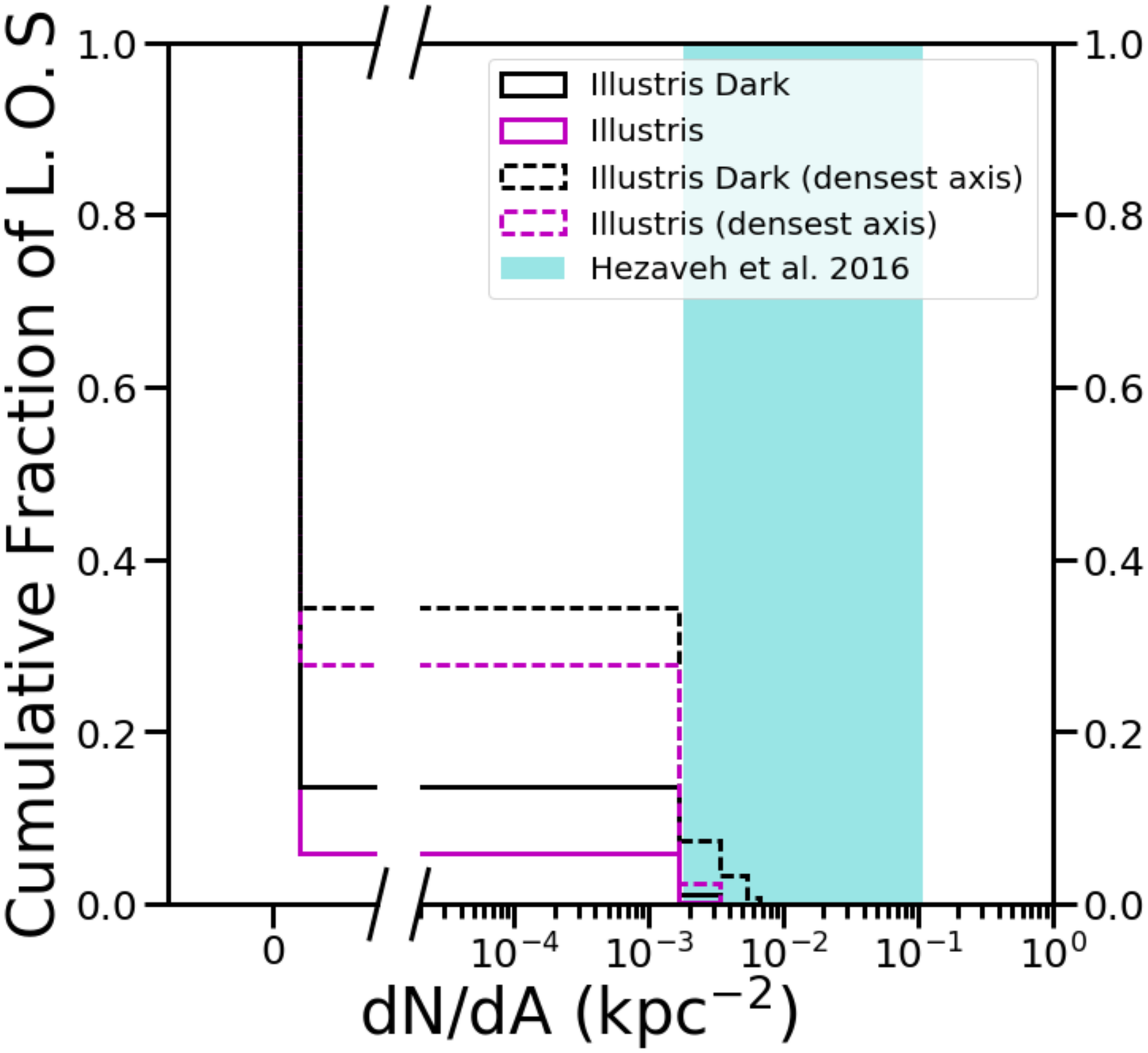}
	\centering
	\caption[projected ]{Subhalo surface number density distributions.  Left: Fraction of lines of sight through our Illustris (magenta) and Illustris Dark (black) halos that contain a given number of subhalos per unit area.  The bin on the far left corresponds to sight lines with no substructure.  We count only subhalos with $M_{\rm halo} > 10^9$ M$_\odot$ in cylinders of radius $R = 10$ kpc, which is typical (though somewhat larger) than an Einstein radius for strong lenses of this type.    The solid lines show results when each halo is viewed along 100 random lines of sight.  The dashed lines include one sightline per halo, viewed along each host halo's densest axis.   Right: the same distributions shown cumulatively.  The shaded regions in both panels show the substructure surface density derived from \cite{Hezaveh2016} for the same subhalo mass cut.  The simulated halos typically have lower substructure fractions than observed, with many lines of sight containing no subhalos at all.   This is especially true for the full physics simulations.
	}
	\label{fig:Surface1}
\end{figure*}

\begin{figure*}
	\includegraphics[width=3.2in, height=3.2 in, trim = 0.5in 0 0.5in 0]{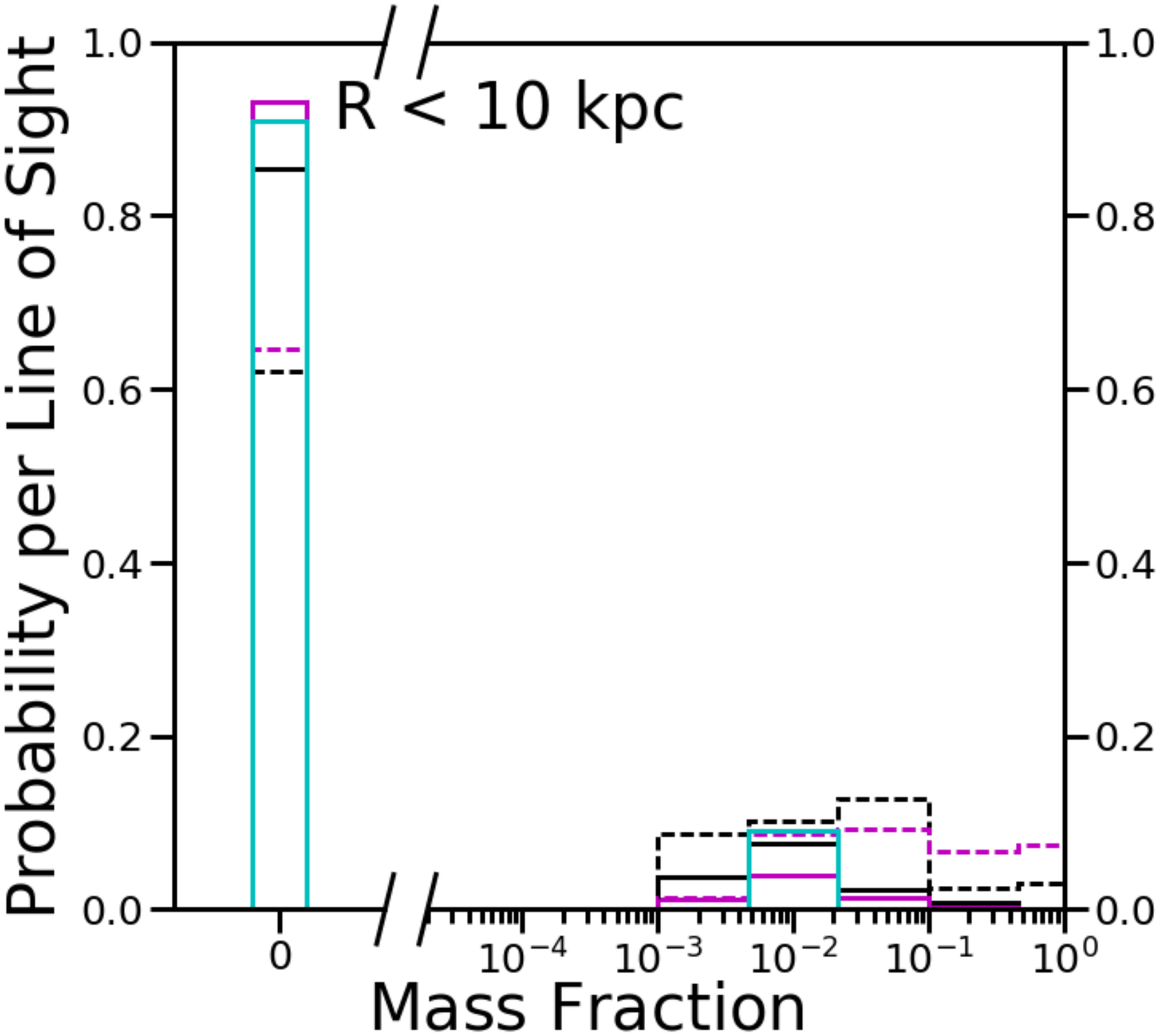}
	\hspace{.2in}
	\includegraphics[width=3.2 in, height=3.2 in, trim = 0.5in 0 0.5in 0]{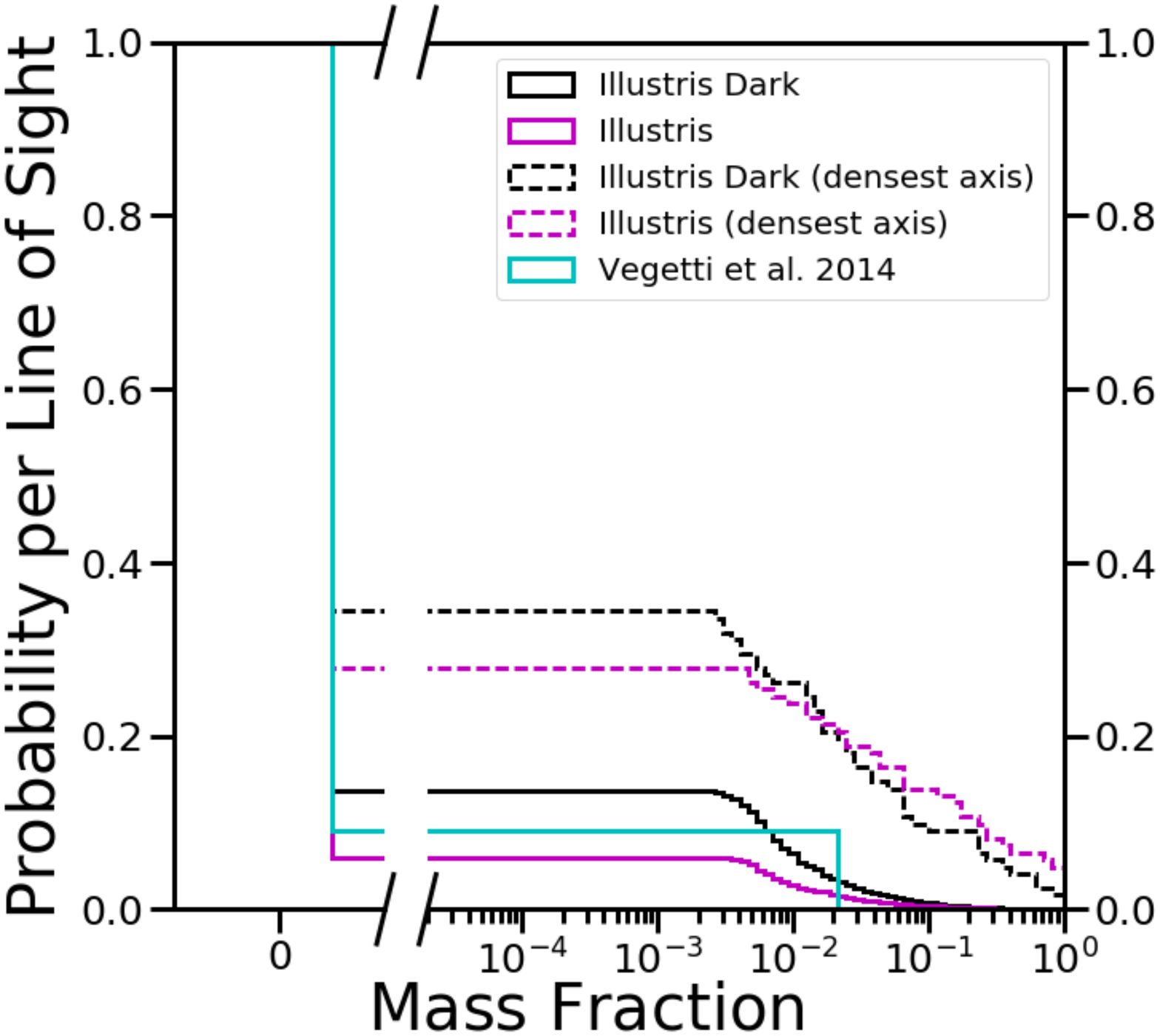}
	\centering
	\caption[substructure mass fractions]{Subhalo mass fraction distributions.  Left: Fraction of lines of sight through our Illustris (magenta) and Illustris Dark (black) halos that contain a given  mass fraction in substructure within cylinders of radius $R=10$ kpc with length $L = 2 \times R_{200}$.  Right: the same distributions shown cumulatively. We have counted the mass in subhalos $M_{\rm halo} > 10^9$ M$_\odot$ whose centers sit within the cylinder.  The cyan histogram corresponds to the reported results of \citet{Vegetti2014}. The solid lines show results when each halo is viewed along 100 random lines of sight.  The dashed lines include one sightline per halo, viewed along each host halo's densest axis.}
	\label{fig:massfrac1}
\end{figure*}

\begin{figure*}
	\includegraphics[width=3.2in, height=3.2 in, trim = 0.5in 0 0.5in 0]{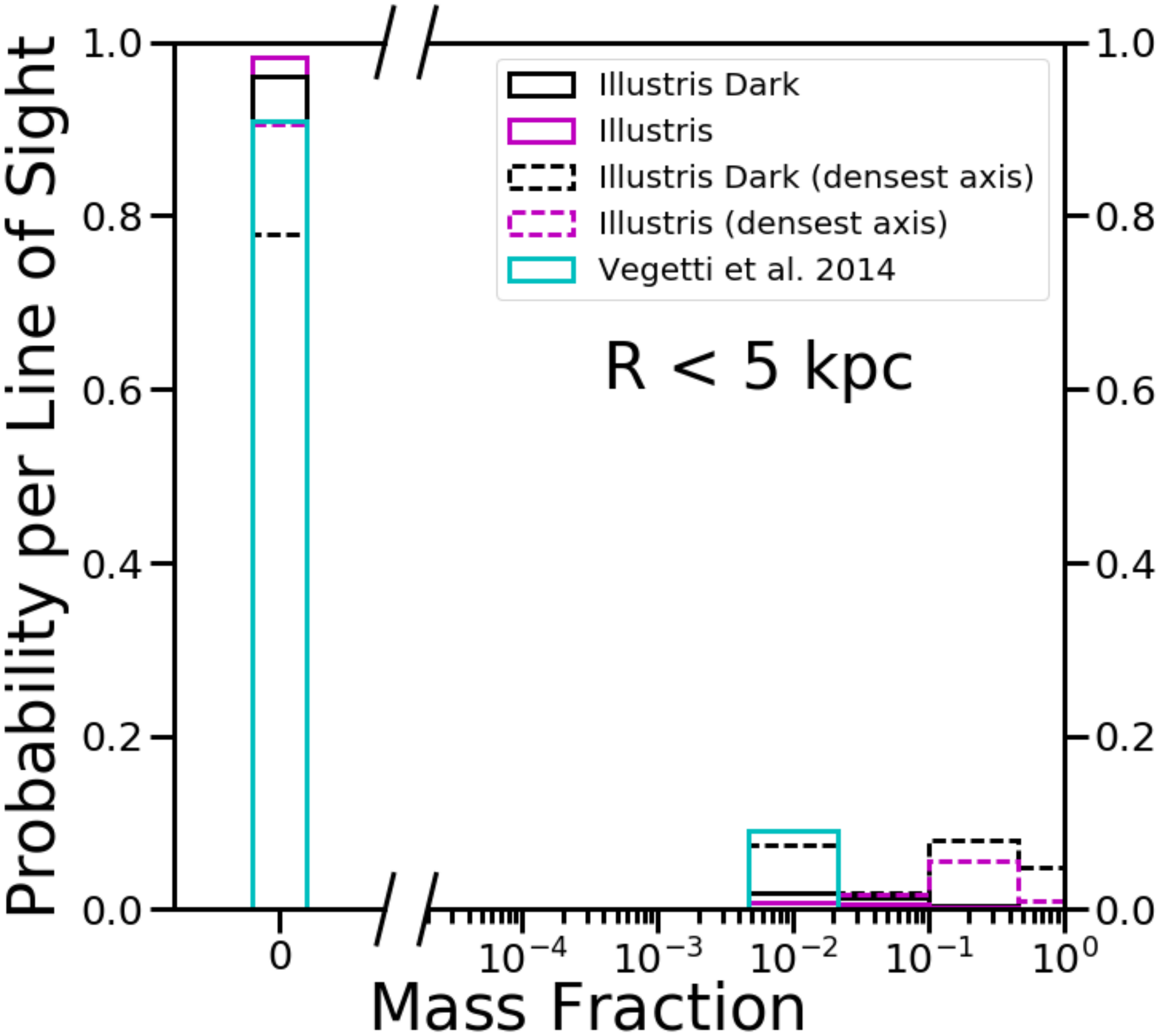}
	\hspace{.2in}
	\includegraphics[width=3.2 in, height=3.2 in, trim = 0.5in 0 0.5in 0]{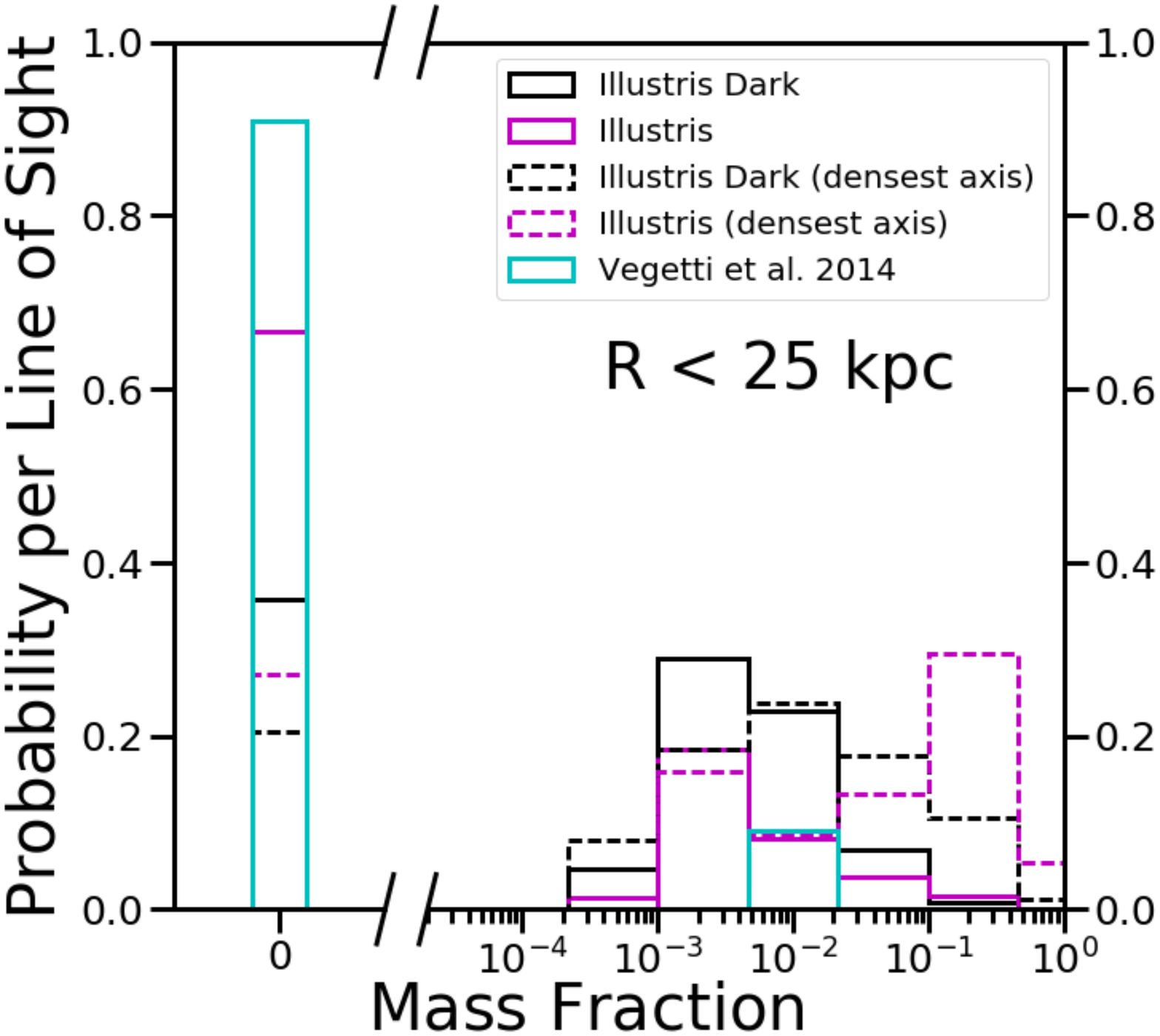}
	\centering
	\caption[projection variation]{The same as figure 5 except varying the assumed projected radius. Predictably, changing the radius to 5 kpc drastically lowers the lines of sight with a detected substructure, while increasing to 25 kpc increases it. }
	\label{fig:varyR}
\end{figure*}

\subsection{Implications for subhalo lensing}
	
Gravitational lensing studies are sensitive to the projected mass along the line of sight near the lens plane of the host.  Substructure constraints are often quoted in terms of the projected mass fraction in substructure \citep[e.g.][]{Dalal02,Vegetti2014} or the projected subhalo mass function \citep{Hezaveh2016}.  We explore predictions for both approaches here.

Figure \ref{fig:Surface1} provides a comparison to the results of  \citet{Hezaveh2016}, who report projected subhalo counts per unit area based on their lensing analysis.  For each host halo in our Illustris and Illustris Dark sample, we construct 100 random lines of sight and count the number of subhalos $N$ more massive than $M_{\rm halo} = 10^9 \msun$ within a projected cylinder of radius $R=10$ kpc and length $L = 2 \times R_{200}$ through the halo.  The fraction of lines of sight that have a given subhalo count per unit area is shown as a histogram on the left and cumulatively on the right, with solid black lines corresponding to the DMO simulation and solid magenta lines corresponding to the hydrodynamical simulation.  We see that the vast majority of sight lines in the Illustris sample ($\sim 95\%$) include no subhalos above our mass cut within this projected radius.  The fraction of empty sight lines drops only slightly to $\sim 85\%$ in the Illustris Dark sample. 

The cyan bands show the reported projected subhalo count from \cite{Hezaveh2016}, where the width of the band represents the 95\% confidence interval \footnote{Hezaveh et al. report a differential mass function per unit subhalo mass.  We have derived the gray band by integrating their mass function to $10^{9}$ $M_{\odot}$, assuming their reported  power law.}.  The \cite{Hezaveh2016} range is noticeably offset from the distribution seen in Illustris, with more substructure than would be expected along a typical sightline.  Only $\sim 0.25\%$ ($\sim 1\%$) of the slight lines from Illustris (Illustris Dark) are consistent with the reported counts.   One possible reason for this is that a halo is more likely to be a strong lens if it is viewed along its major axis towards the observer.  The dashed lines in Figure \ref{fig:Surface1} shows results if we restrict each halo to be viewed only along its densest axis.  With this restriction, the fraction of sight lines consistent with the \cite{Hezaveh2016} range increases to $\sim 2.5\%$ ($7\%$) and the fraction of empty cylinders drops to $\sim 65 \%$ in Illustris and $\sim 72\%$ in Illustris Dark. It is important to note that while the vast majority of our sightlines produce no detections, the comparison to the Hezaveh range is not an ideal one given the observational range was based on a single detection, additionally the \citet{Hezaveh2016} system is at a higher redshift $\it{z}$ = 0.299, and thus the system could not have had enough time to destroy a significant amount of substructure.  We now turn to the larger comparison set of \cite{Vegetti2014}.

Figure \ref{fig:massfrac1} shows a similar analysis to that presented in Figure \ref{fig:Surface1},  now recast for comparison to the substructure mass fraction results reported in the gravitational imaging analysis of 11 systems reported in \cite{Vegetti2014}. We use the same procedure as outlined in the compairison to \citet{Hezaveh2016}, however now instead of only counting the number of subhalos detected above $10^9$ M$_\odot$ we explicitly account for their masses. In order to compute the mass fractions in the simulated halos, we include the total mass of all subhalos with centers that lie within the cylinder and divide by the mass of all particles within the cylinder. This means that there are cases where some fraction of the subhalo mass that sits outside the cylinder's radius is included, but the approach is reasonable because the centers of halos are most important in producing lensing anomalies. The line colors and types mirror those in Figure \ref{fig:Surface1} and the cyan histogram shows the distribution from the \citet{Vegetti2014} sample given that they include 11 systems with one detection with a mass fraction of 0.0215 \citep{Vegetti2010}.  It is important to note that the non-detections depend on the lowest subhalo mass that can be measured, so as techniques and observations improve, the space below mass fractions of $\simeq$ $10^{-3}$ will be filled in with detections of smaller halos both in the current sample of lenses that can be measured and in lower mass hosts or less dense lines of sight with lensing efficiencies that are too weak provide a currently detectable signal. 

With this fiducial comparison, we see that the distribution from the \citet{Vegetti2014} is intermediate between those derived from Illustris and Illustris Dark, though it is consistent with both within the measured uncertainty.   Most of the lenses in the Vegettie sample ($\sim 90\%$) include no detected substructure, as expected from the simulations.  Interestingly, the fraction of empty sight lines is predicted to be lower than observed if we restrict ourselves to the densest axis projections.

There also appears to be mismatch in that the simulations show a tail of very high mass fraction projections.  This  appears to be partially an artifact of our adding the entire subhalo mass in every subhalo mass fraction estimate in cases where large subhalos sit at the edge of the cylinder.  To test this we compute the mass fraction two alternative ways.   First, we estimate the total subhalo mass by counting every particle associated with a subhalo that was within the cylinder. This cuts down on adding a significant amount of mass that is outside the cylinder, but with the added problem that often we are counting very low density parts of halos that are outside of the cylinder, which will probably not contribute to the lensing signal. This method reduces the number of non-detections; however, most of these non-detections end up in a tail down to low mass fractions caused by only the outskirts of a halo being within the cylinder, and are most likely not dense enough to cause a detectable lensing perturbation. The second method is similar, but excludes particles from subhalos that do not have centers within the cylinder.  With this method, the mass fraction of systems does decrease by about a factor of 2.  

The surface density of subhalos (and subhalo mass fractions) that we expect to see along a given line of sight through a halo is highly sensitive to the projected radius used.  This expectation can be seen in Figure \ref{fig:radial}, where we have shown that subhalos tend to be evacuated near a halo's center owing to tidal destruction.  Though surface counts can include halos with large 3D radius that just happen to project within small radii, the effect is still important.  Figure \ref{fig:varyR} reproduces the analysis shown in in Figure \ref{fig:massfrac1} but with cylinders of both smaller ($R=5$ kpc) and larger ($25$ kpc) radii.  We see that the fraction of sight lines with zero subhalos increases dramatically for the 5 kpc case and decreases for the 25 kpc cylinder.   Typical Einstein ring radii for massive galaxy lenses are  $\sim 5$ kpc \citep{Bolton2008}, so in using $R=10$ kpc in the discussion above we are biasing our results towards somewhat higher substructure counts than might be expected in observations.  The $R=25$ kpc projections clearly result in much higher mass fractions and fewer empty sight lines than observed, but this is to be expected given the mismatch between $R$ and typical Einstein ring sizes.  Interestingly, the $R=5$ kpc distribution from Illustris 1 along the densest axis is in better agreement with the Vegetti et al. sample than was the $R=10$ kpc case,  as might be expected.  However, the limited force and mass resolution of the Illustris simulations subject the $5$ kpc projections to potentially under-predicting substructure at small radius due to over merging.    This is why we have used the 10 kpc projections as our fiducial comparisons above. 

In summary, given the limited size of the data comparison sample, the predicted substructure fractions seen in both Illustris and Illustris dark appear to be consistent with what has been observed by \citet{Vegetti2014}.  As sample sizes increase, we would expect observations to more closely align with the predictions of full hydrodynamic simulations if LCDM is indeed the correct underlying model.

\section{Conclusion} \label{s:conclusion}

In this paper, we have used the Illustris simulations \citep{Nelson2015,Vogelsberger2014} to investigate the dark substructure content of dark matter halos chosen to be representative of those studied in substructure lensing analyses ($M_{\rm halo} = 0.5-4 \times 10^{13}$ M$_\odot$; see Figure \ref{fig:one}) and explored how substructure content changes from the dark-matter-only run of Illustris Dark to the full physics run of Illustris.   Our sample consists of more than 100 such host halos at $z=0.2$ and we use them to make projections for substructure lensing studies, specifically for the detection of substructures more massive than $10^9$ M$_\odot$, comparable to reported detections \citep{Vegetti2014,Hezaveh2016}.

We find that the addition of full hydrodynamics reduces the over all subhalo mass function within the virial radius of typical lens-host halos by about a factor of two (Figure \ref{fig:hmf}), an effect that appears to be driven by enhanced central destruction caused by the additional central host galaxy potential (Figure \ref{fig:radial}).    
Naively one would expect this to translate to a factor of two decrease in the expected substructure mass fraction to be observed in lensing studies.  However, when viewed in projection to mimic the lensing signal, the large halo-to-halo scatter leads to a more nuanced prediction (Figures \ref{fig:Surface1}-\ref{fig:varyR}).
Specifically, most of lines of sight through projected cylinders of size close to an Einstein radius contain no substructure at all.  For cylinders of radius $R=10$ kpc, the fraction of empty sight lines rises from $\sim 85\%$ to $\sim 95\%$ as we go from dark matter only simulations to full physics simulations.   This large number of expected non-detections implies that in order to constrain CDM or probe the lower limit on galaxy formation by detecting dark subhalos, we may need hundreds of lensing systems as well as predictions that include the effects of central galaxies.

We note that this is not the first time that a difference between dark matter only and baryon physics simulations has been pointed out at the mass scales relevant for subhalo lensing. \cite{Despali2016} used the EAGLE simulation and the Illustris full hydrodynamics run to show a similar offset. This effect appears to continue to the present day with mostly the same offset as shown by \cite{Chua16}. Additionally, several results on the scale the Milky Way show a similar offset between simulations with and without galaxy formation included \citep{Wetzel2016,Zhu2016,Sawala2016,SGK2017}.  All of these works see subhalo destruction when comparing baryonic and dark matter only simulations at the factor of two level within the virial radius.  The main driver of this effect at the Milky Way scale appears to be an additional tidal field introduced by adding a physical galaxy to the system. Indeed, the effect can be reproduced by simply adding in the potential of a galaxy to a dark matter only simulation \citep{SGK2017}.

As we move towards an era where we expect many more strong lensing systems suitable for substructure searches, we will need to take into account potential biases in lensing probability.  Specifically, if lens systems are biased to be viewed along dense lines of sight, we find that this can reduce the number of non-detections from $\sim$85-95\% in the case of random lines of sight to $\sim$60-85\% in the case of looking along the most dense axis of a halo (depending on the projected radius). It is important to note that while the number of suitable lenses for substructure explorations is rather small today,  we expect the number to explode in the near future owing to large-area, deep surveys such as DES, LSST, and Euclid. Simultaneously, our ability to detect substructure will substantially improve as more studies are done with ALMA, and in the near future with JWST. We expect the ability of substructure lensing to discriminate between dark matter theories will dramatically increase in the near future from two complementary approaches.  

\section*{Acknowledgments} \label{s:Acknowledgements}

We thank the Illustris collaboration for providing their simulation data, and for making all of their data publicly available and easy to use.  We also thank Manoj Kaplinghat, Davide Fiacconi, and Fangzhou Jiang for helpful discussions.  AGS was supported by an AGEP-GRS supplement to NSF grant AST- 1009973.   JSB acknowledges support from NSF grants AST-1518291 and PHY-1520921, NASA grants NNG16P128C, HST-AR-13888.003-A, HST-AR-14282.001-A, and HST-AR-14554 from the Space Telescope Science Institute, which is operated by AURA, Inc., under NASA contract NAS5-26555. 
MBK acknowledges support from NSF grant AST-1517226 and from NASA grants NNX17AG29G and HST-AR-13888, HST-GO-14191, HST-AR-14282, and HST-AR-14554 from the Space Telescope Science Institute, which is operated by AURA, Inc., under NASA contract NAS5-26555. AMN acknowledges support from the UCI Chancellor's ADVANCE Postdoctoral Fellowship.

\appendix

\section{Ambiguity in subhalo masses} \label{s:AppA}

Halo finders assign masses to subhalos in different ways, and therefore produce different subhalo mass functions even when applied to the same simulations.   We illustrate this problem by analyzing our Illustris Dark halo sample using three different halo finders.  In addition to the default Illustris halo finder SUBFIND \citep{Springel01}, we use Rockstar \citep{Behroozi13} and Amiga Halo Finder \citep[AHF; ][]{Knollmann09}. Each of these these halo finders work in different ways, and we refer the reader to the specific papers  for in depth explanations of their differences.  We provide a short description of each method below.

Rockstar works by running a 3D FoF algorithm to identify overdensities, and then a 6D phase space algorithm is run to identify halos, the linking length is adaptively tuned so that some fraction (by default 70\%) of the particles in that group are linked together with at leasts one other particle. Seed halos are placed at the bottom of those groups with particles assigned to their closest seed. Finally, unbinding is performed, and then subhalo properties are calculated. In AHF, dark matter structure and substructure is identified via a grid, where the grid is refined iteratively based on the local density. In practice, if there are more particles in a cell than a number which can be set by hand, the cell is divided in half and the process is repeated. Once this procedure is finished, the most dense areas are called the halo centers and particles are assigned to those halo centers.  Finally, unbound particles are removed iteratively, with particles moving at some multiple of the escape velocity (by default 1.5$v_{esc}$) removed. In contrast to these, the default Illustris halo finder is SUBFIND. SUBFIND is based on a hierarchical Friends of Friends (FoF) algorithm, where the FoF linking length is reduced in discrete steps. Within the FoF group, a smoothing kernel is used to estimate the density within the group. Any locally overdense area of the FoF group is labeled a “substructure candidate”, and once the algorithm has identified all locally overdense regions, it assigns dark matter particles to those candidates, and undergoes an unbinding procedure eliminating unbound particles. If the remaining candidate has enough particles in it to pass some pre set threshold (32 in Illustris), it is counted as a subhalo.

\begin{figure*}
	\includegraphics[width=6.0in, height=3.0 in, trim = .2in 1.2in 0 1.2in]{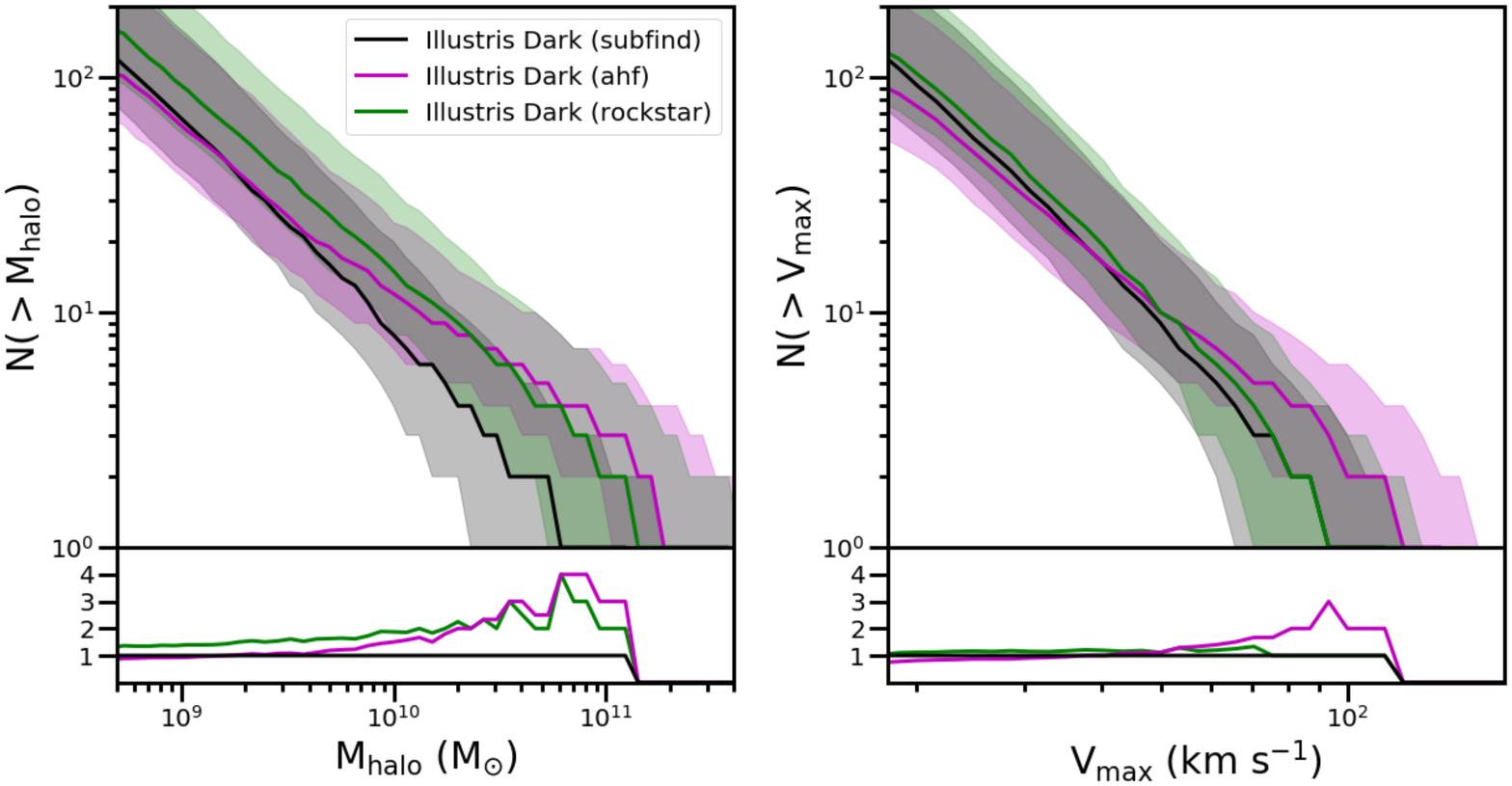}
	\centering
	\caption[halo finders fid]{The subhalo mass functions (left) and subhalo $V_{max}$ functions (right) for the galaxies in our sample compared between different halo finders (AHF and ROCKSTAR) run with their fiducial parameters.}
	\label{fig:A1}
\end{figure*}

\begin{figure*}
	\includegraphics[width=6.0in, height=3.0 in, trim = .2in 1.2in 0 1.2in]{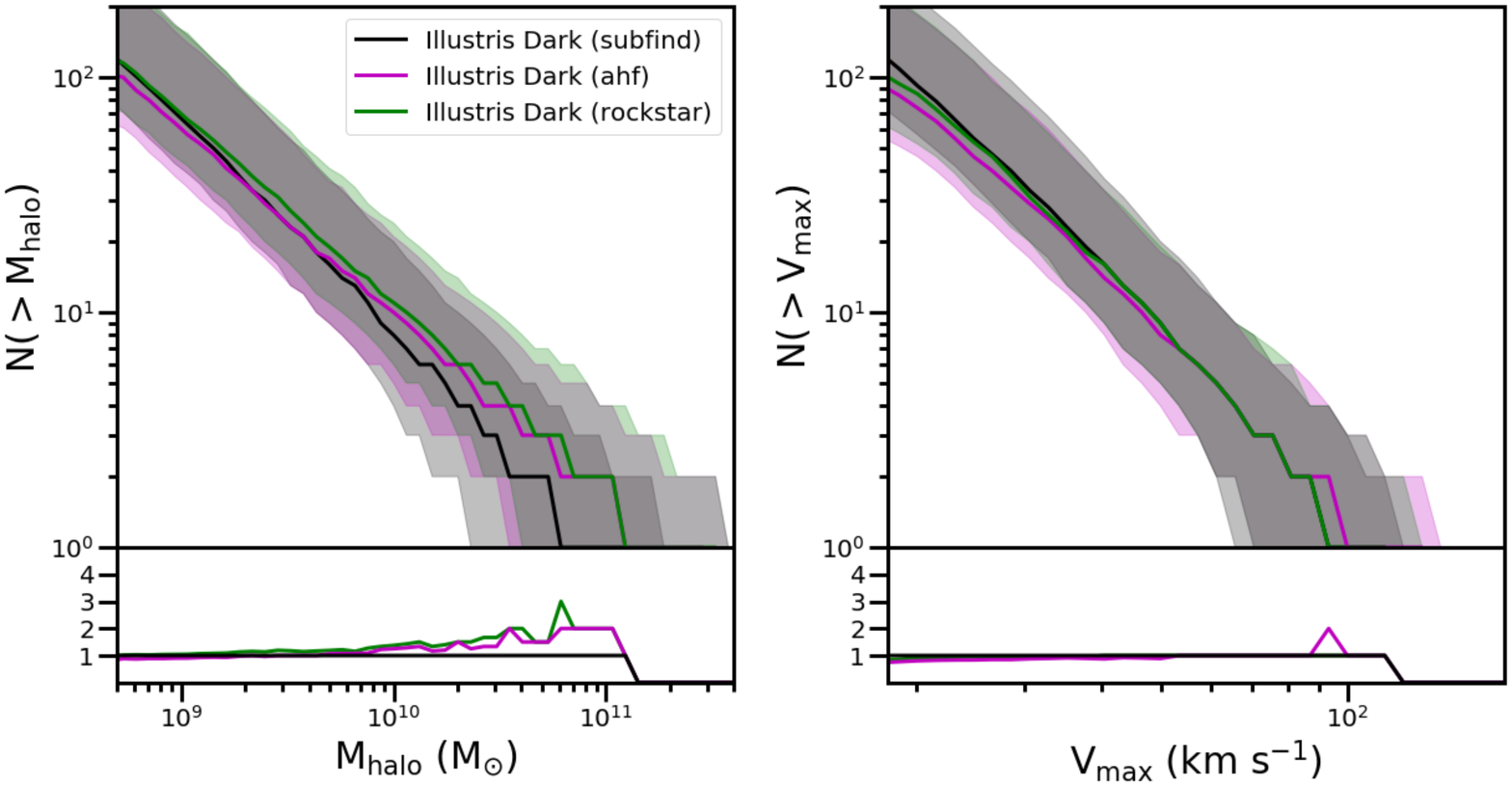}
	\centering
	\caption[halo finders mod]{Same as in figure A1; however, the fiducial parameters for ROCKSTAR and AHF have been modified. First, the velocity at which particles are determined to be unbound in AHF was lowered to 1.25 $v_{esc}$ (from the default 1.5 $v_{esc}$) and the binding criteria for rejection in Rockstar was modified to 95\% (from the default 50\%).}
	\label{fig:A2}
\end{figure*}

The subhalo mass functions are shown in the left hand panel of Figure \ref{fig:A1}. In general, the fiducial Illustris halo finder (SUBFIND) has the lowest abundance of halos, and Rockstar the highest, with AHF somewhat in the middle, although the halo mass function produced by AHF seems to have a different slope. This disagreement can be due to two competing reasons; either the halo finders are finding different populations of halos, or they are finding the same populations of halos but assigning them different masses. We argue that the significant disagreement can be almost completely attributed to assignment of particles to halos.  This can be illustrated by looking the associated $V_{max}$ function (right panel of Figure \ref{fig:A1}). While the agreement is not perfect, it is significantly better than in the halo mass functions and this implies that it is mainly a mass definition issue, and not that the halo finders are identifying different populations of halos.  

The differences in the mass assignments can be attributed to different ways in which the halo finders assign bound particles.  AHF and SUBFIND use a 3D algorithm with an iterative unbinding criteria. Rockstar calculates the fraction of particles that are bound to the halo, and if it is within some acceptable tolerance (by default 50$\%$ bound fraction), the halo is accepted; if not it is thrown out. To test this idea, we can modify the binding criteria of AHF and Rockstar.  This is illustrated in Figure \ref{fig:A2}. Here we set the AHF binding criteria to reject particles moving faster than 1.25$v_{esc}$ (where the the default 1.5$v_{esc}$).  We have modified Rockstar to reject halos that have more than 95$\%$ of their particles unbound (raised from the default 50$\%$).  Once this is done, we see much better agreement between the three halo finders.  

The above exercise illustrates that subhalo mass is a fairly subjective measure and therefore not an ideal choice for direct comparisons subhalo lensing studies. When subhalos are counted using their $V_{max}$ values, the results are in much better agreement between halo finders.
 This is because $V_{max}$ probes the inner regions of the halo, and is therefore less sensitive to bound versus unbound particles. As the field moves towards more precise comparisons,  $V_{max}$, or similarly the mass within some small radius, will be a robust parameter with which to compare simulations and observations. This is similar to what is advocated by \cite{Minor2016}. 
 
 For a more thorough analysis of the difficulties involved in halo finding, particularly in defining halo masses, we refer the reader to \cite{Knebe11}, who carried out a much broader comparison of halo finders and echo the sentiment that a parameter such a $V_{max}$ should be used in place of halo mass.

\section{Baryon Dominated Substructures} \label{s:AppB}

It is possible that baryonic clumps rather than dark matter clumps will be detected in strong lensing searches for substructure.  Concern for such a scenario was raised by \cite{Gilman2017} and \cite{Hsueh2017b}, who showed that luminous substructures can cause anomalous lensing signals in theory, and potentially confirmed by \cite{Hsueh2017a}, who found a flux ratio anomaly from an edge-on disk.  

\begin{figure}
	\includegraphics[width=3.5in, height=3.0 in, trim = 1.0in 0 0 0]{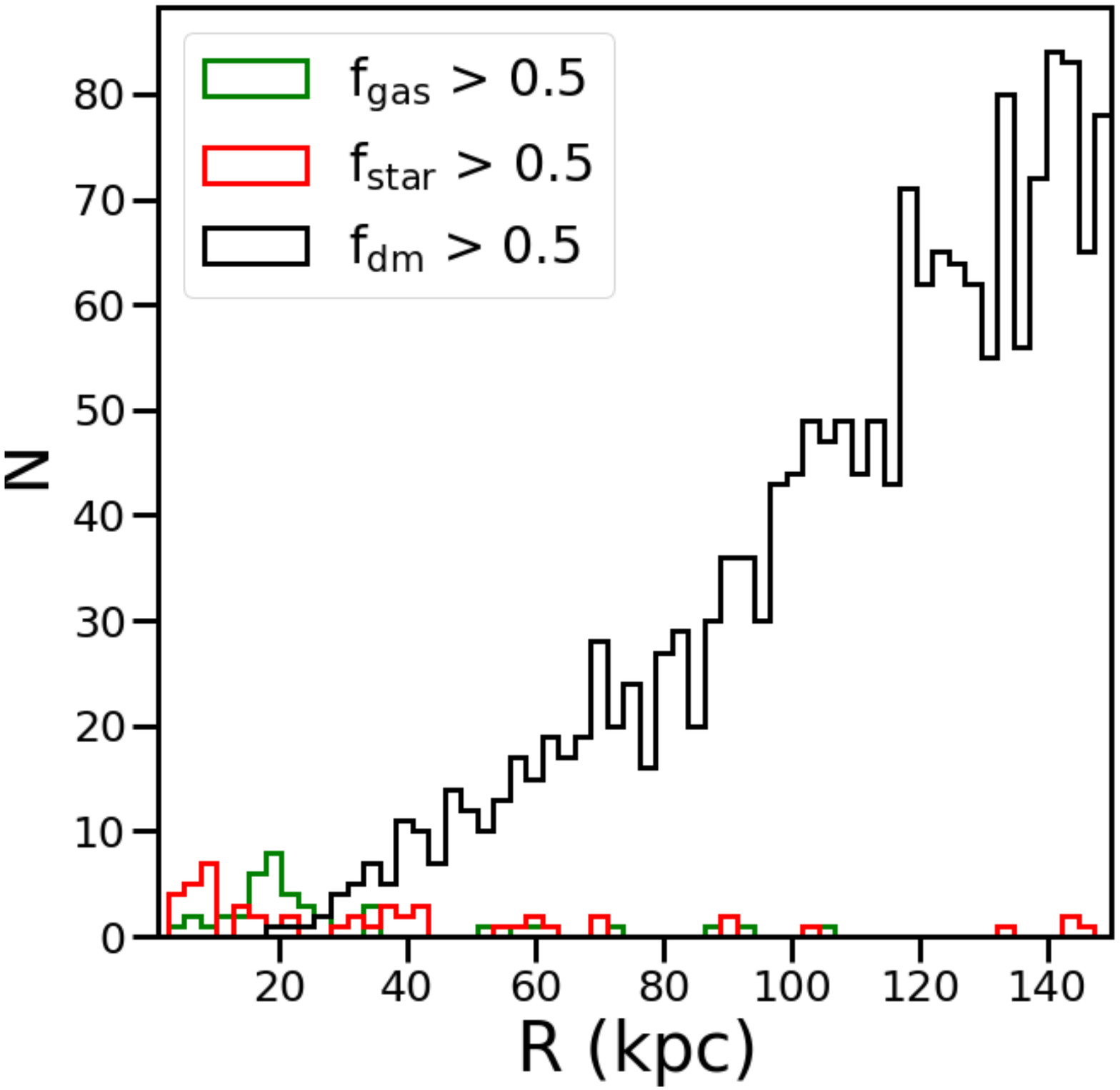}
	\centering
	\caption[particle types]{Objects from the SUBFIND subhalo catalog within 150 kpc of all of the systems selected from Illustris above $10^{9}$ $M_{\odot}$ in halo mass. The histograms are colored by the most dominant particle in the object, red if the stellar mass is over 50\% the total mass, green if the gas mass is over 50\% of the total mass, and black if the halo is dark matter dominated.}
	\label{fig:app}
\end{figure}

As mentioned in Section 2, there are several baryon-dominated structures listed as subhalos in the Illustris catalogs among our sample of 122 galaxy hosts.  Figure \ref{fig:app} shows the radial distribution of all of the subhalos in our fiducial sample stacked together.  The green and red histograms show gas-dominated and star-dominated systems, respectively,  while the black histogram shows dark-matter dominated subhalos.  There are about 100 baryon-dominated "subhalos" out of $\sim 10,000$ subhalos in total.  Most of them exist within the central $\sim 20$ kpc of the galaxy. We have chosen to remove these systems from our analysis, which focuses on predictions for {\em dark} substructure.  If the majority of these baryonic substructures are ``real" predictions -- as apposed to spurious numerical artifacts associated with halo finding in the vicinity the galaxies themselves --  they would need to be treated as foregrounds in any dark subhalo search.

\bibliographystyle{mn2e}
\bibliography{lensing_paper_refs.bib}

\label{lastpage}
\end{document}